# Can Free-Response Questions Be Approximated by Multiple-Choice Equivalents?

Shih-Yin Lin and Chandralekha Singh

*Department of Physics and Astronomy, University of Pittsburgh, PA, 15260*

Abstract

We discuss a study to evaluate the extent to which free-response questions can be approximated by multiple-choice equivalents. Two carefully designed research-based multiple-choice questions were transformed into a free-response format and administered on the final exam in a calculus-based introductory physics course. The original multiple-choice questions were administered in another, similar introductory physics course on the final exam. Our findings suggest that carefully designed multiple-choice questions can reflect the relative performance on the free-response questions while maintaining the benefits of ease of grading and quantitative analysis, especially if the different choices in the multiple-choice questions are weighted to reflect the different levels of understanding that students display.

## I. INTRODUCTION

When it comes to assessing students' learning in physics, there is always a concern about the format of the assessment tool. While a multiple-choice (MC) test provides an efficient tool for assessment because it is easy to grade, instructors are often apprehensive to use such a tool because a free-response format facilitates a more accurate understanding of students' thought processes. In addition, free-response questions allow students to get partial credit for displaying different levels of understanding, which is appreciated by many instructors and students. Thus, there appears to be a trade-off between the two assessment tools. If instructors choose to implement a multiple-choice test, they often feel that they are completely sacrificing the benefits that the free-response questions could provide.

There have been previous efforts devoted to investigating whether a multiple-choice assessment and a free-response assessment can yield similar results in assessing students' ability in physics. For example, in Hudson and Hudson's study[1] students' performances on three multiple-choice exams were compared to their performance on a total of 14 (hand graded) free-response "pop-test" problems given over the semester. A good correlation between the multiple-choice exams and the free-response problems was found, especially as the number of MC questions increased. Scott, Stelzer, and Gladding also reported a high consistency when students' ranking based on the multiple-choice exam scores was compared to the ranking based on students' written explanations graded by the instructors in an introductory electricity and magnetism course.[2] These studies provide evidence that good MC questions in physics can give an assessment result qualitatively similar to that obtained from the administration of free-response (FR) questions.

Indeed, many of the well-known tests designed to evaluate students' learning and difficulties in physics were presented in MC format.[3-13] As indicated by the research literature, the difficulties students have related to a given topic can be classified into relatively few categories. If the choices in the MC questions are designed carefully to reflect the common difficulties students have, it is possible that the multiple choice questions will faithfully reflect the performance on the equivalent free-response questions while maintaining their benefits of ease of grading and comparison of classes taught using different instructional approaches. Here, we present a study designed to investigate the relation between students' performance on quantitative free-response questions and equivalent questions in MC formats.

In particular, we converted two research-based MC questions into free-response format and administered them on the final exam in a calculus-based introductory physics course (course A). The original MC questions were administered in an equivalent introductory physics course (course B) on the final exam. Students' performance in two different courses was compared. Moreover, we investigate the correlation between students' actual performance on the free-response questions and a "simulated" multiple-choice performance had the problems been given in the MC format in course A.

## II. METHODOLOGY

Figures 1 and 2 present the MC questions that were administered in this study. Question 1 concerns an object at rest on an inclined plane. Students were asked to find the magnitude of the frictional force acting on the object, which is equal to $mg\sin\theta$ (the component of the gravitational force parallel to the inclined plane, where $m$ is the mass of the object, $g$ is the gravitational field strength, and $\theta$ is the inclination angle) according to Newton's 2nd Law. Research[14] suggests that many students struggle with this question because they believe that the magnitude of static friction ($f_s$) is always equal to its maximum value, the coefficient of static friction ($\mu_s$) times the normal force ($F_N$). This notion is not valid for this question because the maximum value of static friction is larger than the actual frictional force needed to hold the object at rest. Other student difficulties found from a pilot study include the confusion between static and kinetic friction and the challenge in decomposing the force correctly. The alternative choices in the MC question are designed to reflect these difficulties. Table I presents the choices in algebraic form where all symbols have their standard meanings. In order to compare students' performance in the two courses and avoid cueing the students who were provided the MC questions, the actual choices presented were not symbolic but numerical.

Question 2 concerns a roller coaster cart on a frictionless track. The question asks for the normal force ($F_N$) acting on the cart when it goes over a hump, which can be obtained using the principles of conservation of mechanical energy and Newton's 2$^{nd}$ law in the non-equilibrium situation (with centripetal acceleration involved). Previous research[15] indicates that a common difficulty introductory physics students have is that they think of a non-equilibrium situation involving centripetal acceleration as an equilibrium situation by treating the centripetal force as an additional force. The correct use of the centripetal acceleration and Newton's 2$^{nd}$ Law in this question should yield $F_N - mg = -\frac{mv_B^2}{r}$ (where $v_B$ is the speed of the cart at the point of interest and $r$ is the radius of the circle at point B). However, students who treat it as an equilibrium question and believe that the centripetal force is an additional force obtain an answer of the type $F_N - mg - \frac{mv_B^2}{r} = 0 \Rightarrow F_N = mg + \frac{mv_B^2}{r}$, which has an incorrect sign. In addition to this common mistake, a pilot study indicates that some students incorrectly believe that the normal force is equal to the gravitational force ($F_N = mg$) without contemplating the centripetal acceleration. On the other hand, there are also students who completely skip the gravitational force and claim that $F_N = \frac{mv_B^2}{r}$. Moreover, some students have difficulty figuring out the speed of the cart at point B because they are confused by the two different heights provided. These common difficulties are incorporated in the design of the multiple-choice questions.

Table II presents the choices in algebraic form where all symbols have their standard meaning.

The MC questions and the corresponding free-response questions were administered on the final exam in two introductory physics courses (with 185 and 153 students, respectively.) The Force Concept Inventory[13] implemented both at the beginning and the end of the semester indicates that students in these two courses are comparable. As shown in Table III, there is no statistically significant difference between the students in the two courses in terms of their FCI scores. The free-response questions administered in one course were the same as their counterparts in the MC format administered in the other course, except that there were no choices provided to the students. Students' performance on the free-response questions was graded using rubrics, with a full score of 1 for each question. Tables IV and Table V present the rubrics for each question, respectively. The rubrics were developed by two researchers who took into account the common difficulties found on the free-response problems. Different partial scores were assigned to the students based on their problem solving approach (Question 1) or the principles used (Question 2). The common mistakes students made and the corresponding points taken off are also listed in Tables IV and V.

To construct two types of "simulated" MC scores from the answers students provided for the free-response questions, student responses were first binned into different categories by comparing and matching their answers to the different choices in the MC questions. For the dichotomous MC simulation, a score of 1 (correct choice) or 0 (incorrect choice) was then assigned for the various categories. For the weighted MC simulation we assigned partial credit to different binned responses based on approaches students used for the free-response questions; this was done to simulate the partial credit that is usually awarded for a free-response question. The scores assigned to each of the choices in this "weighted" MC simulation are shown in Table I and

Table II. The different weights for the choices are commensurate with the rubrics used to grade the free-response questions.

To summarize, students who were given the free-response questions were graded using three different methods: a rubric, a dichotomous MC simulation, and a weighted MC simulation. On the other hand, students who were given the MC questions were graded using two methods involving dichotomous or weighted scoring. Table VI summarizes the different methods used to analyze student performance in the two courses.

**III. FINDINGS**

Tables VII and VIII present the percentage of students who were binned into different categories by matching their free-response answers (in course A) to the choices in the corresponding MC Questions 1 and 2, respectively. The percentage of students binned into different categories based on their multiple-choice answers in course B is also listed. We find that out of the 153 students in course A, 84% and 88% of their free-response answers could be matched to the *a priori* choices in the multiple-choice Questions 1 and 2, respectively. The findings suggest that a carefully designed, research-based multiple-choice question can reasonably reflect the distribution of common difficulties students have as detected in their free-response answers.

Students' average performance on Questions 1 and 2 is presented in Fig. 3. The black and white labels are used to distinguish students in course B, who were given the multiple-choice questions, and students in course A, who were given the free-response questions. In general we find that the trends for students' performance in the two courses are similar regardless of the question format they were given. For example, comparing students' rubric-graded free-response performance in course A to students' dichotomous multiple-choice performance in course B, we find that in both courses, students who displayed a higher level of expertise on the final exam (e.g., students in groups 4 and 5) on average typically scored higher on Questions 1 and 2 in both formats than those who did not perform as well on the final exam overall.

Comparing students' performance in course A to course B, as shown in Fig. 3, we also find that there is a better correspondence between students' performance on free-response questions in one class and the multiple-choice questions in another class if partial credit is awarded for both types of questions.

The reason free-response performance for students in one class has a better match with the weighted multiple-choice performance in the other class (compared to the dichotomous multiple-choice performance) is that the weights for the weighted MC performance were similar to those used in the rubric to score the free-response questions.

Table IX presents the correlation coefficients between students' free-response performance (graded using the rubrics) and the simulated multiple-choice performance (both dichotomous and weighted) in course A. It shows that the correlation coefficient is always higher for weighted multiple-choice simulation than the dichotomous multiple-choice simulation. As one might expect, the correlation coefficient between the rubric-graded FR performance and the *weighted* multiple-choice simulation is very high for both Questions (0.925 and 0.945 for Question 1 and Question 2, respectively.) When it comes to the correlation between the rubric-graded FR performance and the *dichotomous* multiple-choice simulation, the correlation coefficient for Question 1 is similarly high but the correlation coefficient for Question 2 drops to 0.483. This result indicates that the correlation between free-response and simulated *dichotomous* MC performance is higher for the question with a single, very-strong distracter choice (Question 1) in MC compared to the question with several distracter choices (Question 2), each of which represents a different level of understanding. On the other hand, if the multiple-choice simulation is graded in a *weighted* manner there is not much difference between the question with a single strong distracter and the question with multiple appealing distracters. The finding re-emphasizes that it is beneficial to grade the MC questions by assigning different weights to the alternative choices.

While the comparisons discussed above suggest that the MC questions can reasonably well reflect students' performance on FR question, especially when the MC questions are graded in a weighted manner, we note that there are some improvements that can be made to the current MC questions to improve their quality further. For example, in our current version of the multiple-choice questions the mistake of using 1-D kinematics equations instead of the principle of conservation of mechanical energy to find the speed at point B in Question 2 cannot be detected in the multiple-choice version because both methods yield the same numerical value for an option in the MC question. This deficiency may be remedied in the future by adding another question that asks students to select the physics principles they believe are applicable for solving the problem. A similar strategy of adding a companion question can also be used in Question 1 as students' FR answers suggest that two different problem solving approaches—(1) correctly using $\sum F_x = f_s - mg_x = 0$ (where $mg_x$ represents the component of the gravitational force parallel to the plane) but confusing $\cos\theta$ with $\sin\theta$, or (2) treating $f_s$ as equal to the normal force—can both result in the answer $f_s = mg\cos\theta$ (presented in choice (d) in the multiple choice question), while the latter approach contains a more serious error than the former. When grading students' free-response answers using a rubric, the former approach is assigned a maximum score of 0.9 while the latter approach is assigned a maximum score of 0.2 (for finding the normal force correctly). By adding a companion question that asks students to also find the normal force acting on the object and comparing students' answers for the frictional force to that of the normal force, the weight for choice (d) can be assigned more precisely to reflect the students' understanding of the subject matter tested. In addition, we would like to note that there is a very close correspondence between the MC questions and the FR questions used in this study. As indicated from previous research, students' responses to similar physics problems can be context-dependent.[16] Even if two problems involve the same underlying physics concepts, different contexts or representations may trigger different responses.[16-18] In our study students in both assessment formats were asked to solve the same quantitative physics questions; in other words, the contexts and the presentations of the questions are the same except that the numerical choices provided in the multiple-choice format are removed from the free-response format. It is possible that if there is more variation between the MC and the corresponding FR questions, the good correlations exhibited in our study may not be found.

**IV. CONCLUSIONS**

In this study we compare students' performance on two research-based, quantitative multiple-choice questions (in which the distracter choices correspond to students' common difficulties) and their free-response equivalents (in which all the choices are removed). We find that the trends in student performance on the research-based multiple-choice questions given to one class and the free-response questions given to another equivalent class are similar in that those who displayed a higher level of expertise on the final exam in each of the classes performed better on the questions than those who displayed a lower level of expertise, regardless of the format of questions provided to them. Moreover, there is a good match between students' free-response answers in one class and the *a priori* choices in the MC questions administered to another class.

The findings suggest that research-based MC questions can reasonably reflect the relative performance of students on the free-response questions, especially if the answers for the MC questions are graded in a weighted manner by assigning partial credit to the different choices. We note that, similar to the rubrics for the free-response questions, the weightings for the different alternative choices in the MC questions reflect the fact that some mistakes are not as "bad" as others despite the fact that they lead to students selecting an incorrect choice. If different scores are assigned to the different choices in the MC questions in the weighted model to reflect the different levels of understanding students display, there is a good similarity found between students' MC performance in one class and the free-response performance in another class.

Although the similarity found between students' MC performance and FR performance suggests that the MC questions can reasonably reflect students' performance on the FR questions, it is not our intention to argue that research-based MC exams possess all the advantages of free response exams and that the former should replace the latter. If an instructor would like to get detailed information about some particular student's complete problem solving process, or if the problem involves complicated mathematical computation for which an instructor would like to evaluate students' abilities to correctly apply the mathematical tools/models in detail, a free-response question may be more suitable. On the other hand, if an instructor is more concerned about students' average performance as a whole, and if there are pragmatic concerns for grading, our study indicates that research-based MC questions can be a good choice. Also, free-response questions are useful only if they are graded carefully based on a good rubric. When they are graded leniently without a good rubric, the resulting scores will not typically reflect the appropriate level of student understanding. On the other hand, a computer can grade weighted MC questions with weights corresponding to a good rubric for each distracter choice. Once the weights for the choices have been determined via research, MC questions can be as accurate for assessment purposes as rubric-based free-response questions without the time constraint.

We re-emphasize that the fidelity of a MC question to a free-response performance depends strongly on the incorrect choices given.[19] If students' common difficulties found via research are incorporated, instructors can utilize MC questions without sacrificing accuracy in assessing students' thinking processes. It will be of great benefit to the physics education community if more research-based MC questions with appropriate weights for different choices are developed and made available to the community. Future research can focus on developing more research-based MC questions that target different topics in physics and refining the qualities of the questions through careful testing and evaluation (e.g., by comparing students' performance on the FR questions to the MC questions developed.)

12

Table I. The algebraic form for the choices in Question 1 and the different scores assigned in the "weighted multiple-choice" simulation (the correct answer is indicated by the shaded background). In order to avoid cueing the students who were provided with the MC questions, the actual choices were not symbolic but numerical.

| Choice | Question 1 | Score |
|---|---|---|
| (a) | $mg \sin \theta$ | 1.0 |
| (b) | $\mu_k mg \cos \theta$ | 0.3 |
| (c) | $\mu_s mg \cos \theta$ | 0.5 |
| (d) | $mg \cos \theta$ | 0.9 |
| (e) | None of the above | 0.0 |

Table II. The algebraic form for the choices in Question 2 and the different scores assigned in the "weighted multiple-choice" simulation (the correct answer is indicated by the shaded background). Except for choice (d), the speed at point B ($v_B$) is calculated correctly using the square root of $2gh_1$ in choices (b), (c) and (e).

| Choice | Question 2 | Score |
|---|---|---|
| (a) | $F_N = mg$ | 0.2 |
| (b) | $F_N = mg + m\dfrac{v_B^2}{r}$ | 0.8 |
| (c) | $F_N = m\dfrac{v_B^2}{r}$ | 0.7 |
| (d) | $F_N = mg - m\dfrac{v_B^2}{r}$<br>$v_B$ calculated using $\sqrt{2g(h_1 + h_2)}$ | 0.9 |
| (e) | $F_N = mg - m\dfrac{v_B^2}{r}$ | 1.0 |

Table III. Students' average score on the Force Concept Inventory (FCI) implemented at the beginning (pre) and the end (post) of the semester. The $p$-values computed using one-way ANOVA are also presented.

|  | Course A (N=153) | Course B (N=185) | $p$-value |
|---|---|---|---|
| FCI Pre | 15.10 | 16.32 | 0.071 |
| FCI Post | 19.85 | 20.98 | 0.082 |

Table IV. Summary of rubric for Question 1.

| Problem solving approach | Maximum score | Common mistakes (Points taken off) |
|---|---|---|
| Using $\sum F = 0$ | 1 | (1) Decomposed the force incorrectly: $f_s - mg \cos\theta = 0$ (-0.1) <br> (2) Decomposed the force incorrectly: $f_s - mg/\sin\theta = 0$ (-0.2) <br> (3) Confused weight with mass and multiplied the weight by an additional $g = 9.8$ m/s² (-0.1) |
| $f_s = \mu_s F_N$ | 0.5 | (1) Decomposed the normal force incorrectly (-0.1) <br> (2) $F_N = mg$ (-0.2) <br> (3) Confused weight with mass and multiplied the weight by an additional $g = 9.8$ m/s² (-0.1) |
| $f_k = \mu_k F_N$ | 0.3 or 0.4 [a] | |
| Combined $\mu_s$ and $\mu_k$ | 0.2 | |

[a] If the student lists both $f_s = \mu_s F_N$ and $f_k = \mu_k F_N$ without indicating which is his/her final answer, the maximum score is 0.4. Otherwise, the maximum score is 0.3.

Table V. Summary of rubric for Question 2.

| Description | Correct answer | Common mistakes | Points taken off |
|---|---|---|---|
| Invoking and applying the principle of conservation of mechanical energy to find the speed (0.3 point) | $mg \Delta h = \dfrac{1}{2} m v_B^2$ | use 1-D kinematics equations to find $v_B$ | 0.2 |
| | | incorrect $\Delta h$ | 0.1 |
| Identifying the centripetal acceleration and using Newton's 2nd Law to find the tension (0.7 point) | $a = a_c = \dfrac{v_B^2}{r}$ | $a = 0$, $F_N = mg$ | 0.5 |
| | $F_N - mg = ma_c = -m\dfrac{v_B^2}{r}$ | $F_N = m\dfrac{v_B^2}{r}$ | 0.3 |
| | $F_N = mg - m\dfrac{v_B^2}{r}$ | $F_N = mg + m\dfrac{v_B^2}{r}$ | 0.2 |

Table VI. Summary of grading methods in the two courses.

|  | Course A (given free-response questions, $N=153$) | Course B (given multiple-choice questions, $N=185$) |
|---|---|---|
| Graded using a rubric | Yes | -- |
| Multiple-choice (dichotomous) | simulated [a] | Yes |
| Multiple-choice (weighted) | simulated [a] | Yes |

[a] Student responses were first binned into different categories by comparing and matching their answers to the different choices in the multiple-choice format and then assigning a score as discussed in the text.

Table VII. Percentage of students binned into different categories for simulated MC by comparing their free-response answers (in course A) to the choices in MC Question 1 (the correct answer for this question is indicated by the shaded background). The percentage of students binned into different categories based on their multiple-choice answers in course B is also listed.

| Choice | Question 1 | Course A (FR) % | Course B (MC) % |
|---|---|---|---|
| (a) | $mg \sin\theta$ | 28 | 34 |
| (b) | $\mu_k mg \cos\theta$ | 7 | 7 |
| (c) | $\mu_s mg \cos\theta$ | 46 | 42 |
| (d) | $mg \cos\theta$ | 3 | 6 |
| (e) | None of the above | 16 [b] | 10 |

[b] Both choice (b) and choice (c) were found in one students' free-response answer in this category.

Table VIII. Percentage of students binned into different categories for simulated MC by comparing their free-response answers (in course A) to the choices in MC Question 2 (the correct answer for this question is indicated by the shaded background). The percentage of students binned into different categories based on their multiple-choice answers in course B is also listed.

| Choice | Question 2 | Course A (FR) % | Course B (MC) % |
|---|---|---|---|
| (a) | $F_N = mg$ | 9 | 29 |
| (b) | $F_N = mg + m\dfrac{v_B^2}{r}$ | 31 | 25 |
| (c) | $F_N = m\dfrac{v_B^2}{r}$ | 28 | 15 |
| (d) | $F_N = mg - m\dfrac{v_B^2}{r}$, $v$ calculated using $\sqrt{2g(h_1 + h_2)}$ | 6 | 3 |
| (e) | $F_N = mg - m\dfrac{v_B^2}{r}$ | 14 | 28 |
| -- | Other | 12 | 1 |

Table IX. Correlation (N=153) between the free-response performance, graded using the rubrics (FR), vs. the simulated multiple-choice performance for Questions (Q) 1 and 2.

| | FR vs. simulated multiple-choice (dichotomous) | | FR vs. simulated multiple-choice (weighted) | |
|---|---|---|---|---|
| | Q 1 | Q 2 | Q 1 | Q 2 |
| Correlation coefficient ($r$) | 0.890 | 0.483 | 0.925 | 0.945 |
| $p$-value | 0.000 | 0.000 | 0.000 | 0.000 |

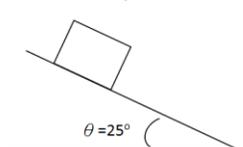

Fig. 1. Problem Statement for Question 1.

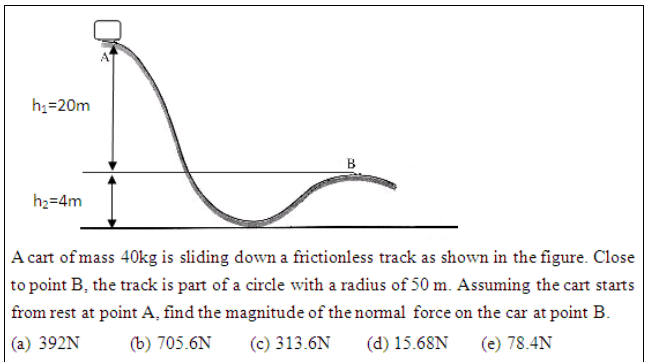

Fig. 2. Problem statement for Question 2.

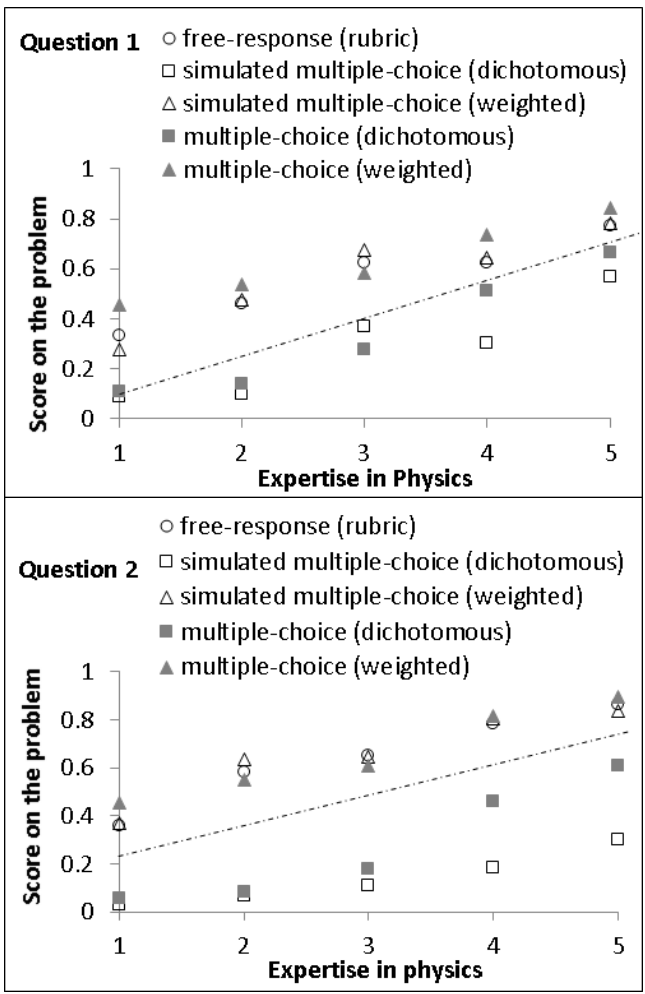

Fig. 3. Students' average performance on (a) Question 1 and (b) Question 2. The white and black data labels are used to indicate students in course A (who were given the free-response question) and students in course B (who were given the multiple-choice questions), respectively. A dashed line is included on the figure to separate the data for the dichotomous case vs. the case where partial credit is assigned to the students.